\documentclass[twocolumn,showpacs,superscriptaddress]{revtex4}
\usepackage{amssymb}
\usepackage{amsmath}
\usepackage{graphicx}
\usepackage{epsfig}
\usepackage{amsfonts}

\begin{document}

\title{ Induced Entanglement Enhanced by Quantum Criticality }
\author{Qing Ai}
\affiliation{Department of Physics, Tsinghua University, Beijing 100084, China}
\author{Tao Shi}
\affiliation{Institute of Theoretical Physics, Chinese Academy of
Sciences, Beijing, 100080, China}
\author{Guilu Long}
\affiliation{Department of Physics, Tsinghua University, Beijing 100084, China}
\affiliation{Tsinghua National Laboratory for Information Science and Technology, Beijing
100084, China}
\author{C. P. Sun}
\affiliation{Institute of Theoretical Physics, Chinese Academy of
Sciences, Beijing, 100080, China}

\begin{abstract}
Two qubit entanglement can be induced by a quantum data bus
interacting with them. In this paper, with the quantum spin chain in
the transverse field as an illustration of quantum data bus, we show
that such induced entanglement can be enhanced by the quantum phase
transition (QPT) of the quantum data bus. We consider two external
spins simultaneously coupled to a transverse field Ising chain. By
adiabatically eliminating the degrees of the chain, the effective
coupling between these two spins are obtained. The matrix elements
of the effective Hamiltonian are expressed in terms of dynamical
structure factor (DSF) of the chain. The DSF is the Fourier
transformation of the Green function of Ising chain and can be
calculated numerically by a method introduced in [O. Derzhko, T.
Krokhmalskii, Phys. Rev. B \textbf{56}, 11659 (1997)]. Since all
characteristics of QPT are embodied in the DSF, the dynamical
evolution of the two external spins displays singularity in the
vicinity of the critical point.
\end{abstract}

 \maketitle

\section{Introduction}

Entanglement lies at the heart of quantum mechanics and thus can be
regarded as a resource for quantum information processing. In recent
years, many people have demonstrated that quantum entanglement can
offer an intrinsic clarification of quantum criticality of a many
body system. For example, in Refs.\cite{Osterloh,Osborne} it was
proved that the derivative of the nearest-neighbor entanglement
diverged at the critical point. Furthermore, one of the authors CPS
and his collaborators studied the dynamical ultrasensitivity of the
induced quantum critical system, which concerned the excited states
as well as the ground state \cite{Quan,Sun}. Usually, the former has
not been discussed in the investigation for the above mentioned
intrinsic entanglement in many body systems.

In this paper, we consider the quantum entanglement of the many body
problem in the point view of quantum information processing, whether
or not quantum criticality of the many body system (as a quantum
data bus) can enhance the entanglement of the external qubits
interacting with the quantum data bus. Actually, the phase
transition indeed creates some entanglement. A case in point is the
superconducting phenomenon \cite{Schrieffer,Tinkham}. As shown in
the BCS theory \cite{Bardeen1,Bardeen2}, a Cooper pair is created
when a conductor transits from a normal state to a superconducting
state. In such a pair, two electrons form a correlated entire over
hundreds of nanometers, which is a long distance in the microscopic
world. We remark that the phase transition of superconducting
happens at finite temperature but this paper will focus on the
occurrence of quantum phase transition (QPT) at zero temperature.

QPT is of critical importance to the quantum statistical physics.
Generally speaking, QPT takes place at zero temperature. It's the
situation where only uncertainty principle plays the major role
while the fluctuation due to the finite temperature does not. As
some parameter is varied, a qualitative change occurs in the ground
state of a quantum many-body system due to QPT
\cite{Sachdev,Sondhi}. At the critical point, long range correlation
also develops in the ground state. This long range correlation
intuitively exhibits greater quantum entanglement between two
points. Refs.\cite{Osterloh,Osborne} indeed showed the entanglement
became singularly longer at the critical point. These considerations
directly motivate us to study the problem.

On the other hand, the interesting thing of this paper is its
meaning for the detection of QPT by coupling the quantum critical
system to the external detector-two qubits. And various efforts have
been devoted to this field. In Ref.\cite{Quan}, when an external
spin underwent a transition from a pure state to a mixed state, the
decay of the Loschmidt echo of its coupling environment described by
a transverse field Ising model (TFIM) was greatly enhanced. Others
suggested two external spins to detect QPT by simultaneously
coupling to an XY model environment \cite{Yi,Yuan}. The former
researches showed the critical phenomena of QPT with exact solvable
models. However, they mainly focused on $\sigma^z\sigma^z$
interaction between the external spins and the the Ising chain.
Thus, by extending $z$-$z$ coupling to a more general coupling form,
we explore the possibility of detecting QPT by two central spins.

The rest of paper is organized as follows. In the next section, we
describe the model as two external spins coupled to a transverse
field Ising chain. The calculation of the effective Hamiltonian
between the two spins is outlined using Fr\"{o}hlich transformation
in Sec. III. The matrix elements of the effective Hamiltonian are
given in terms of dynamical structure factor (DSF). In Sec. IV, the
DSF is numerically calculated and the relation between the coupling
constants of two spins and the variable parameter is given. Finally,
the significant results are concluded in Sec. V.

\section{Model Description}

\begin{figure}[ptb]
\includegraphics[bb=132 275 455 572,width=5 cm]{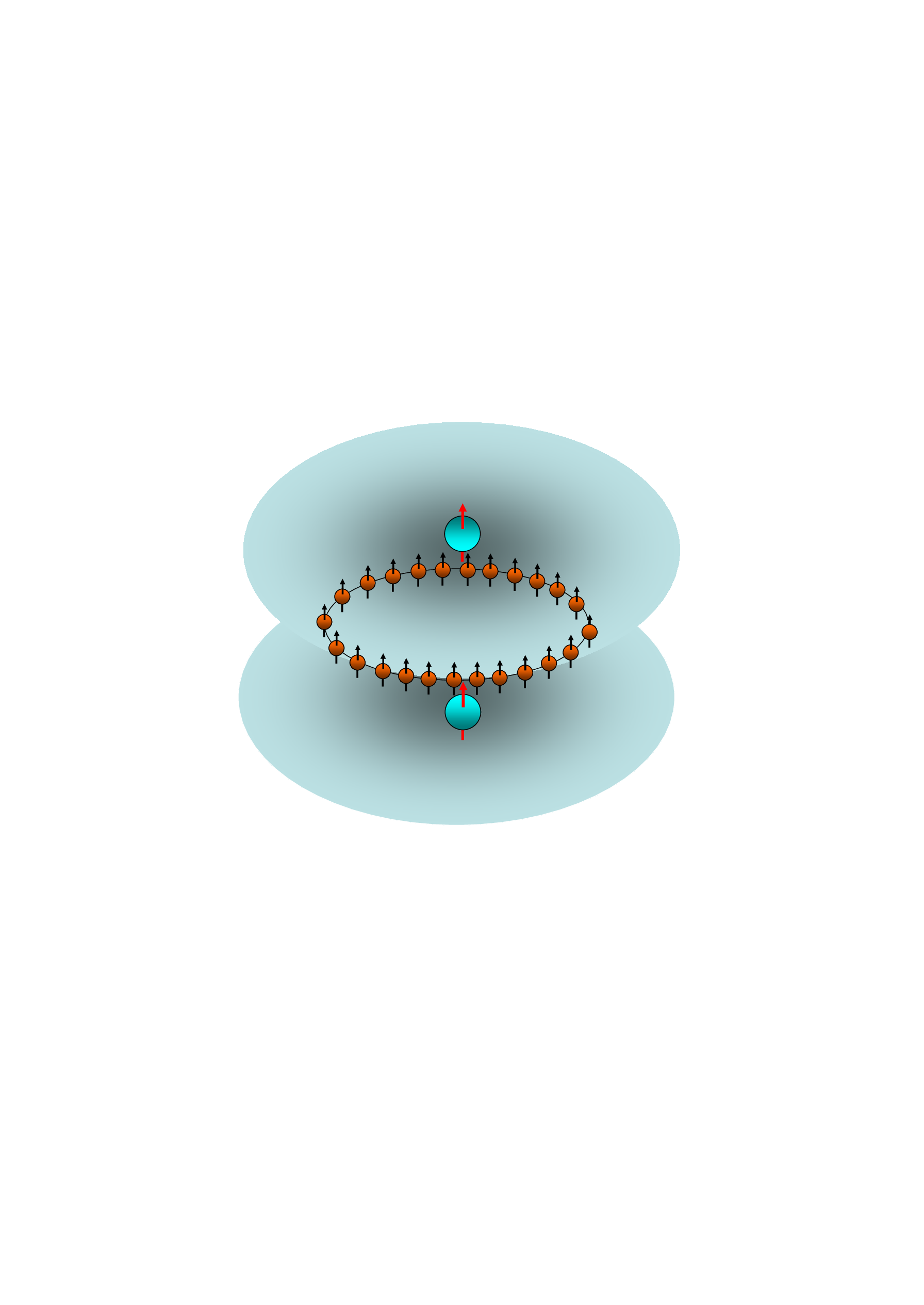}
\caption{(color online) Schematic diagram of two external spins (C)
simultaneously coupling to a 1-D Ising chain (E).} \label{Sketch}
\end{figure}

Two external spins are simultaneously coupled to an environment
described by a one dimensional transverse field Ising chain. By
exchanging spin angular momentum with the chain, the two
noninteracting spins attain effective interaction between them. To
study the dynamic detail we consider the system depicted in
Fig.\ref{Sketch}. The Hamiltonian of this model reads
\begin{equation}
H =H_0+H_I= H_C+H_E+H_I \text{,} \label{originalH}
\end{equation}
where
\begin{equation}
H_{C} =\frac{\mu }{2}(\sigma _{A}^{z}+\sigma _{B}^{z}) \label{HC}
\end{equation}
is the unperturbed Hamiltonian for the external spins,
\begin{equation}
 H_{E} =-\sum\limits_{j=1}^{N}(\Gamma \sigma
_{j}^{z}+J\sigma _{j}^{x}\sigma _{j+1}^{x}) \label{HE}
\end{equation}
is the Hamiltonian of the transverse field Ising model, and
\begin{align}
H_{I}
&=\sum\limits_{\alpha=A,B}\sum\limits_{j=1}^{N}\frac{J_{\alpha}}{\sqrt{N}}(\sigma
_{j}^{x}\sigma _{\alpha}^{x}+\sigma _{j}^{y}\sigma
_{\alpha}^{y}) \notag \\
&=\sum\limits_{\alpha=A,B}\sum\limits_{j=1}^{N}\frac{2J_{\alpha}}{\sqrt{N}}(\sigma
_{j}^{+}\sigma _{\alpha}^{-}+\sigma _{j}^{-}\sigma _{\alpha}^{+})
 \label{HI}
\end{align}
is the interaction between the external spins and the environment.
Here $\sigma_{\alpha}^{\beta}$ and $\sigma_{j}^{\beta}$
($\alpha=A,B,$ $\beta=x,y,z$) are Pauli operators for the two
external spins and the Ising chain respectively,
$\sigma_{\alpha,j}^{\pm}=(\sigma_{\alpha,j}^{x}+i\sigma_{\alpha,j}^{y})/2$
are the corresponding raising and lowering operators,
$J_{\alpha}/\sqrt{N}$ is the homogeneous coupling constants between
$\alpha$th external spin and $j$th site of Ising chain with $N$
being the number of sites in the chain.

First of all, $H_E$ is diagonalized with the combination of the
Jordan-Wigner transformation \cite{Jordan}
\begin{equation}
c_j=\exp(\pi
i\sum\limits_{k=1}^{j-1}\sigma_k^+\sigma_k^-)\sigma_j^- \text{,} \\
\end{equation}
and the Bogliubov transformation \cite{Pfeuty}
\begin{equation}
\eta_k=\frac{1}{2}\sum\limits_{j=1}^{N}[(\phi_{kj}+\psi_{kj})c_j+(\phi_{kj}-\psi_{kj})c_j^+)]
\text{,}
\end{equation}
where for $\lambda=J/\Gamma\neq1$,
\begin{eqnarray}
\phi_{kj}&=&\sqrt{\frac{2}{N}}\sin(kj)\text{ for }k>0 \text{,} \\
\phi_{kj}&=&\sqrt{\frac{2}{N}}\cos(kj)\text{ for }k\leq 0 \text{,} \\
\psi_{kj}&=&-\frac{1}{\Lambda_k}[(1+\lambda\cos
k)\phi_{kj}+\lambda\sin k\phi_{-kj}] \text{.}
\end{eqnarray}
Here,
\begin{equation}
\Lambda_{k}=\sqrt{1+\lambda^2+2\lambda\cos k}
\end{equation}
is the energy spectrum of the quasi particle with $k=2\pi m/N$,
$m=-N/2,\cdots,N/2-1 \text{ for } \text{even } N$, and
$m=-(N-1)/2,\cdots,(N-1)/2 \text{ for } \text{odd } N$. For
$\lambda=1$ and $m=-N/2$,
\begin{eqnarray}
\Lambda_{k}=0\text{, } \phi_{kj}=\sqrt{\frac{1}{N}} \text{,
}\psi_{kj}=\pm\sqrt{\frac{1}{N}} \text{.}
\end{eqnarray}
Thus, in the quasi-particle representation the Hamiltonian of the
TFIM is rewritten as
\begin{eqnarray}
H_E=2\Gamma\sum\limits_{k}\Lambda_k(\eta_k^+\eta_k-\frac{1}{2})\text{,}
\end{eqnarray}
with corresponding eigen state and energy being
\begin{eqnarray}
\left\vert m\right\rangle&=&\prod_k(\eta_k^+)^{n_k}\left\vert
0\right\rangle \text{,} \\
E_m&=&2\Gamma\sum\limits_{k}\Lambda_k
n_k-\Gamma\sum\limits_{k}\Lambda_k \text{,}
\end{eqnarray}
respectively. Here, $\left\vert 0\right\rangle$ is the ground state
and $n_k=\eta_k^+\eta_k$ is the particle number operator.

\section{Effective Hamiltonian}

Generally speaking, Fr\"{o}hlich transformation
\cite{Frohlich,Nakajima} is widely used in condense matter physics.
It can solve a class of problems such as the induced effective
interaction between two electrons by exchanging phonons with the
crystal lattice. In this paper, the one-dimensional Ising chain
plays the role as a medium to induce the effective interaction
between the two external spins. Therefore, by virtue of Fr\"{o}hlich
transformation, we obtain the effective Hamiltonian between the two
external spins by tracing over the degrees of the environment.

With an appropriate anti-Hermitian transformation $S$ defined by the
matrix elements
\begin{equation}
\left\langle m\right\vert S\left\vert
n\right\rangle=\frac{\left\langle m\right\vert H_I\left\vert
n\right\rangle}{E_n-E_m} \text{,}
\end{equation} which meets the
condition $H_{I}+[H_{0},S]=0$, the effective Hamiltonian is
approximated to the second order as $H_{eff}=H_{C}+H_{el}$, where
\begin{eqnarray}
H_{el} &=&\frac{1}{2}\left\langle 0\right\vert [H_{I},S]\left\vert
0\right\rangle  \label{Hel} \\
&=&\frac{1}{2}\sum\limits_{m}(\left\langle 0\right\vert
H_{I}\left\vert m\right\rangle \left\langle m\right\vert S\left\vert
0\right\rangle -\left\langle 0\right\vert S\left\vert m\right\rangle
\left\langle m\right\vert H_{I}\left\vert 0\right\rangle ) \notag
\end{eqnarray}
with $\{\left\vert m\right\rangle\}$ and $\{E_m\}$ being the eigen
states and eigen energies of $H_E$ respectively, and $\left\vert
0\right\rangle$ its ground state.

The DSF \cite{Derzhko97}
\begin{equation}
S^{\alpha \beta }(k,\omega ) =\sum\limits_{n=1}^N
e^{ikn}\int\limits_{0}^{\infty }\left\langle 0\right\vert \sigma
_{j}^{\alpha }(t)\sigma _{j+n}^{\beta }\left\vert 0\right\rangle
e^{(i\omega-0^+)t}dt
\end{equation}
($\alpha,\beta=x,y$) is the Fourier transformation of the Green
function for the TFIM. It is calculated as
\begin{eqnarray}
S^{\alpha \beta }(k,\omega ) &=&i\sum\limits_{n=1}^N
e^{ikn}\sum\limits_{m}\frac{\left\langle 0\right\vert \sigma
_{j}^{\alpha }\left\vert m\right\rangle \left\langle m\right\vert
\sigma _{j+n}^{\beta }\left\vert
0\right\rangle }{E_{0}-E_{m}+\omega +i0^{+}} \label{S} \\
&=&i\sum\limits_{n=1}^N e^{ikn}\sum\limits_{m}\left\langle
0\right\vert \sigma _{j}^{\alpha }\left\vert m\right\rangle
\left\langle m\right\vert \sigma _{j+n}^{\beta }\left\vert
0\right\rangle \notag \\
&&\times[\wp(\frac{1}{E_{0}-E_{m}+\omega})
-i\pi\delta(E_{0}-E_{m}+\omega)] \notag
\end{eqnarray}
with $\wp(1/x)$ being the principal value of $1/x$.

The right hand side of Eq.(\ref{Hel}) contains the following terms,
which are expressed in terms of DSF's as
\begin{eqnarray}
&&\sum\limits_{m,j,j^{\prime }} \left\langle 0\right\vert \sigma
_{j}^{+}\left\vert m\right\rangle \left\langle m\right\vert \sigma
_{j^{\prime }}^{+}\left\vert
0\right\rangle\wp (\frac{1}{E_{0}-E_{m}+\omega }) \notag \\
&=&\sum\limits_{m,j,j^{\prime }} \left\langle 0\right\vert \sigma
_{j}^{-}\left\vert m\right\rangle \left\langle m\right\vert \sigma
_{j^{\prime }}^{-}\left\vert
0\right\rangle\wp (\frac{1}{E_{0}-E_{m}+\omega }) \notag \\
&=&\frac{N}{4}\text{Im}[S^{xx}(0,\omega )-S^{yy}(0,\omega )]
\text{,}
\end{eqnarray}
\begin{eqnarray}
&&\sum\limits_{m,j,j^{\prime }}\left\langle 0\right\vert \sigma
_{j}^{+}\left\vert m\right\rangle \left\langle m\right\vert \sigma
_{j^{\prime }}^{-}\left\vert 0\right\rangle \wp
(\frac{1}{E_{0}-E_{m}+\omega
}) \notag \\
&=&\frac{N}{4}\{\text{Im}[S^{xx}(0,\omega )+S^{yy}(0,\omega
)]-2\text{Re}S^{xy}(0,\omega )\} \text{,}
\end{eqnarray}
\begin{eqnarray}
&&\sum\limits_{m,j,j^{\prime }} \left\langle 0\right\vert \sigma
_{j}^{-}\left\vert m\right\rangle \left\langle m\right\vert \sigma
_{j^{\prime }}^{+}\left\vert 0\right\rangle\wp
(\frac{1}{E_{0}-E_{m}+\omega
}) \notag \\
&=&\frac{N}{4}\{\text{Im}[S^{xx}(0,\omega )+S^{yy}(0,\omega
)]+2\text{Re}S^{xy}(0,\omega )\} \text{.}
\end{eqnarray}

Then, we rewrite the effective Hamiltonian as
\begin{eqnarray}
H_{eff} &=&\frac{\mu _{A}}{2}\sigma _{A}^{z}+\frac{\mu
_{B}}{2}\sigma _{B}^{z}+g_1(\sigma _{A}^{+}\sigma _{B}^{-}+\sigma
_{A}^{-}\sigma _{B}^{+})
\notag \\
&& +g_2(\sigma _{A}^{+}\sigma _{B}^{+}+\sigma _{A}^{-}\sigma
_{B}^{-})\label{Heff}
\end{eqnarray}
in terms of DSF's, where \cite{note1}
\begin{eqnarray}
\mu _{A} &=&\mu +J_{A}^{2}\{\text{Im}[S^{xx}(0,\mu )+S^{yy}(0,\mu
)-S^{xx}(0,-\mu ) \notag \\ && -S^{yy}(0,-\mu )]
+2\text{Re}[S^{xy}(0,\mu )+S^{xy}(0,-\mu )]\} \text{,} \notag \\
\mu _{B} &=&\mu +J_{B}^{2}\{\text{Im}[S^{xx}(0,\mu )+S^{yy}(0,\mu
)-S^{xx}(0,-\mu ) \notag \\ && -S^{yy}(0,-\mu )]
+2\text{Re}[S^{xy}(0,\mu )+S^{xy}(0,-\mu )]\} \text{,} \notag \\
g_1 &=&J_{A}J_{B}\{\text{Im}[S^{xx}(0,\mu )+S^{yy}(0,\mu
)+S^{xx}(0,-\mu ) \notag \\&&+S^{yy}(0,-\mu )]
+2\text{Re}[S^{xy}(0,\mu )-S^{xy}(0,-\mu )]\} \text{,} \notag \\
g_2 &=&J_{A}J_{B}\text{Im}[S^{xx}(0,\mu )-S^{yy}(0,\mu )
+S^{xx}(0,-\mu )\notag\\&&-S^{yy}(0,-\mu
)]\text{.}\label{coeffecients}
\end{eqnarray}

In the forthcoming section, by using the numerical method in
Ref.\cite{Derzhko97}, the DSF $S^{\alpha\beta}(k,\omega)$ and thus
the matrix elements of the effective Hamiltonian
(\ref{coeffecients}) are calculated explicitly. For further details
about the diagonalization method of the Ising-like model and the
fast scheme for the calculation of Pfaffian, please refer to
Refs.\cite{Derzhko98,Lieb,Pfeuty,Goupalov,Mattis,Jia}.

\section{Critical Coupling}

In order to calculate the DSF numerically, we shall summarize the
numerical method introduced in Ref.\cite{Derzhko97}. For a spin
chain in a transverse field with open ends, the Hamiltonian is
described as \cite{note2}
\begin{eqnarray}
H_E^\prime=\Omega\sum\limits_{j=1}^{N}\sigma_{j}^{z}+J\sum\limits_{j=1}^{N-1}
\sigma_{j}^{x}\sigma_{j+1}^{x}\text{.}\label{He'}
\end{eqnarray}
After the Jordan-Wigner transformation, the Hamiltonian is
transformed into Fermion representation. Then, it is equivalent to
solving the following eigen problem \cite{Lieb,Mattis},
\begin{eqnarray}
\Phi_k(A-B)(A+B)&=&\Lambda_k^2\Phi_k\text{,}\\
\Psi_k(A+B)(A-B)&=&\Lambda_k^2\Psi_k\text{,}
\end{eqnarray}
where $A$ and $B$ are two $N\times N$ matrixes with their matrix
elements being
$A_{ij}=2\Omega\delta_{ij}+J\delta_{i+1,j}+J\delta_{i-1,j}$ and
$B_{ij}=J\delta_{i+1,j}-J\delta_{i-1,j}$. According to the
Wick-Bloch-de Dominicis theorem, the $x$-$x$ correlation function
can be expressed in the form of the Pfaffian of the $2(2j+n-1)\times
2(2j+n-1)$ antisymmetric matrix constructed from elementary
contractions
\begin{widetext}
\begin{eqnarray}
\langle\sigma_j^x(t)\sigma_{j+n}^x\rangle &=&
\langle\varphi_1^+(t)\varphi_1^-(t)\varphi_2^+(t)\varphi_2^-(t)\cdots
\varphi_{j-1}^+(t)\varphi_{j-1}^-(t)\varphi_{j}^+(t)\varphi_1^+\varphi_1^-\varphi_2^+\varphi_2^-\cdots
\varphi_{j+n-1}^+\varphi_{j+n-1}^-\varphi_{j+n}^+\rangle \notag \\
&=&Pf\setcounter{MaxMatrixCols}{20}
\begin{bmatrix}
0 & \langle\varphi_1^+\varphi_1^-\rangle &
\langle\varphi_1^+\varphi_2^+\rangle & \cdots &
\langle\varphi_1^+(t)\varphi_{j+n}^+\rangle \\
-\langle\varphi_1^+\varphi_1^-\rangle & 0 & \langle\varphi_1^-\varphi_2^+\rangle & \cdots & \langle\varphi_1^-(t)\varphi_{j+n}^+\rangle\\
\vdots & \vdots & \vdots & \cdots & \vdots\\
-\langle\varphi_1^+(t)\varphi_{j+n}^+\rangle & -\langle\varphi_1^-(t)\varphi_{j+n}^+\rangle & -\langle\varphi_2^+(t)\varphi_2^+\rangle & \cdots & 0 \\
\end{bmatrix}\text{,}
\end{eqnarray}
\end{widetext}
where
\begin{equation}
\varphi_{j}^{\pm}=c_j^{+}\pm c_j \end{equation} is the linear
comibination of the Fermion operators, and
\begin{eqnarray}
\left\langle\varphi_{j}^{+}(t)\varphi_{m}^{+}\right\rangle
&=&\sum\limits_{p=1}^{N}\Phi _{pj}\Phi _{pm}e^{-i \Lambda _{p}t} \text{,} \\
\left\langle \varphi_{j}^{+}(t)\varphi_{m}^{-}\right\rangle
&=&\sum\limits_{p=1}^{N}\Phi _{pj}\Psi _{pm}e^{-i \Lambda _{p}t} \text{,} \\
\left\langle \varphi_{j}^{-}(t)\varphi_{m}^{+}\right\rangle
&=&-\sum\limits_{p=1}^{N}\Psi _{pj}\Phi _{pm}e^{-i \Lambda _{p}t} \text{,} \\
\left\langle \varphi_{j}^{-}(t)\varphi_{m}^{-}\right\rangle
&=&-\sum\limits_{p=1}^{N}\Psi _{pj}\Psi _{pm}e^{-i \Lambda _{p}t}
\end{eqnarray}
are the elementary contractions of zero temperature obtained from
the finite temperature counterparts in Refs.
\cite{Derzhko97,Derzhko98}.

The Pfaffian is the square root of the determinant of the
corresponding antisymmetric matrix. A fast computation scheme is
given in Ref.\cite{Jia,Derzhko98}. For an $N\times N$ antisymmetric
matrix
\begin{eqnarray}
X=\setcounter{MaxMatrixCols}{20}
\begin{bmatrix}
A & B \\
-B^T & C \\
\end{bmatrix}
\end{eqnarray}
with the dimensions of $A$, $B$, $C$ being $2\times2$, $2\times
(N-2)$, $(N-2)\times (N-2)$ respectively, the Pfaffian of $X$ can be
computed in the following way. Because of
\begin{eqnarray}
\setcounter{MaxMatrixCols}{20}
\begin{bmatrix}
I_2 & 0 \\
B^TA^{-1} & I_{N-2} \\
\end{bmatrix}
X
\begin{bmatrix}
I_2 & -A^{-1}B \\
0 & I_{N-2} \\
\end{bmatrix}
=
\begin{bmatrix}
A & 0 \\
0 & C+B^TA^{-1}B \\
\end{bmatrix}\notag
\end{eqnarray}
with $I_n$ being $n$ dimensional unit matrix, we have
\begin{eqnarray}
\text{Det}(X)=\text{Det}(A) \text{Det}(C+B^TA^{-1}B)\text{.}
\end{eqnarray}
Since antisymmetric $A$ is of the simple form
\begin{eqnarray}
A=\setcounter{MaxMatrixCols}{20}
\begin{bmatrix}
0 & x_{12} \\
-x_{12} & 0 \\
\end{bmatrix}\text{,}
\end{eqnarray}
the $N-2$ dimensional matrix $C+B^TA^{-1}B$ is also antisymmetric.
The above procedure can be repeated times and times again. And the
original matrix $X$ is decomposed into $N/2$ 2D antisymmetric
matrices. Finally, due to $\text{Pf}(A)=x_{12}$, the Pfaffian of
matrix $X$ will be simply a product of $N/2$ numbers obtained from
those $2\times2$ matrices in the above procedure. As it is not
necessary for $A$ to be a 2D matrix at the upper left corner of $X$,
$A$ can be chosen to be a diagonal block such that $\text{Det}(A)$
is the largest, for the stability of the algorithm.

Furthermore, other DSF's can be calculated according to the relation
between correlation functions, that is
\begin{eqnarray}
\left\langle \sigma _{j}^{x}(t)\sigma _{j+n}^{y}\right\rangle
&=&-\left\langle \sigma _{j}^{y}(t)\sigma _{j+n}^{x}\right\rangle \notag \\
&=&\frac{1}{%
2\Omega }\frac{d}{dt}\left\langle \sigma _{j}^{x}(t)\sigma
_{j+n}^{x}\right\rangle \text{,} \\
\left\langle \sigma _{j}^{y}(t)\sigma _{j+n}^{y}\right\rangle &=&-\frac{1}{%
(2\Omega )^{2}}\frac{d^{2}}{dt^{2}}\left\langle \sigma
_{j}^{x}(t)\sigma _{j+n}^{x}\right\rangle \text{.}
\end{eqnarray}

\begin{figure}[ptb]
\includegraphics[bb=93 265 473 561,width=7 cm]{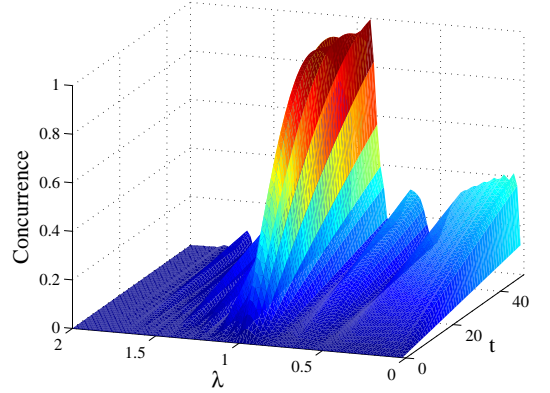}
\caption{(color online) Concurrence evolution vs $\lambda$ for
$N=40$, $\mu=1$, $J_A=J_B=0.1$ with an initial state %
$\left\vert eg\right\rangle$.} \label{concurrence}
\end{figure}

In the last section, we have obtained a typical spin-spin coupling
in the effective Hamiltonian induced by the Ising chain. Driven by
this Hamiltonian, two external spins can be entangled dynamically.
To characterize the extent of entanglement, we use concurrence to
measure the induced entanglement. For an arbitrary state of
two-qubit system described by the density operator $\rho $, a
measure of entanglement can be defined as the concurrence
\cite{Wootters, Wang},
\begin{equation}
C(\rho )=max\{0,\lambda _{1}-\lambda _{2}-\lambda _{3}-\lambda
_{4}\}\text{,}
\end{equation}%
where the $\lambda _{i}$'s are the square roots of the eigenvalues
of the
non-Hermitian matrix $\rho \widetilde{\rho }$ in decreasing order. And%
\begin{equation}
\widetilde{\rho }=(\sigma ^{y}\otimes \sigma ^{y})\rho ^{\ast
}(\sigma ^{y}\otimes \sigma ^{y}),
\end{equation}%
where $\rho ^{\ast }$ is the complex conjugate of $\rho $.

As shown in Fig.\ref{concurrence}, we investigate the evolution of
concurrence under $H_{eff}$. The two external spins start with an
initial product state $\left\vert eg\right\rangle$. As time passes
by, the two spins will eventually evolve into a maximum entangled
state. It can be seen from the figure that the time needed for
reaching maximum entanglement can be greatly shortened in the
vicinity of $\lambda=\Omega/J=1$. In other words, the induced
entanglement between two external spins can be enhanced by quantum
criticality.

\begin{figure}[ptb]
\includegraphics[bb=93 265 473 561,width=7 cm]{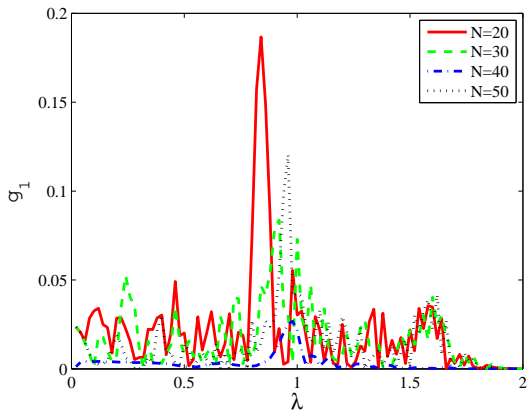}
\caption{(color online) $g_1$ vs $\lambda$ for different $N$'s with
$\mu=1$ and $J_A=J_B=0.1$. Red solid line for $N=20$, Red solid line
for $N=20$, green dashed line for $N=30$, blue dash-dot line line
for $N=40$, black dotted line line for $N=50$.} \label{g1}
\end{figure}

\begin{figure}[ptb]
\includegraphics[bb=100 264 474 561,width=7 cm]{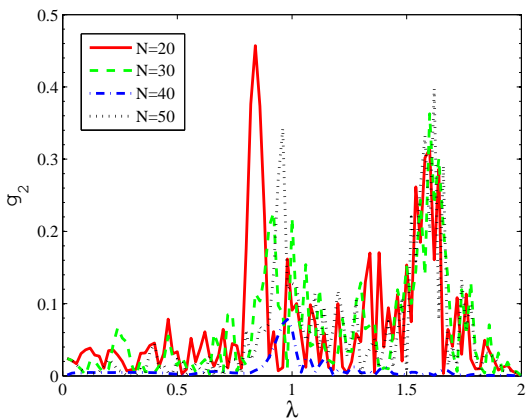}
\caption{(color online) $g_2$ vs $\lambda$ for different $N$'s with
$\mu=1$ and $J_A=J_B=0.1$. Red solid line for $N=20$, Red solid line
for $N=20$, green dashed line for $N=30$, blue dash-dot line line
for $N=40$, black dotted line line for $N=50$.} \label{g2}
\end{figure}

Furthermore, the relation between $g_1$ and $\lambda$ is plotted for
different $N$'s in Fig.\ref{g1}. Although the curves oscillates
shapely, they share a characteristics in common. There is a peak
near the critical point $\lambda=1$ for different curves. It is a
reasonable result since it has been discovered that the entanglement
between two nearest neighbors of the Ising chain achieves maximum
near the critical point \cite{Osterloh,Osborne}. Similar result is
also obtained for the relation between $g_2$ and $\lambda$ in
Fig.\ref{g2}. However, besides the one near $\lambda=1$ there is
another peak around $\lambda=1.5$ except that for $N=40$.

\begin{figure}[ptb]
\includegraphics[bb=95 265 473 562,width=7 cm]{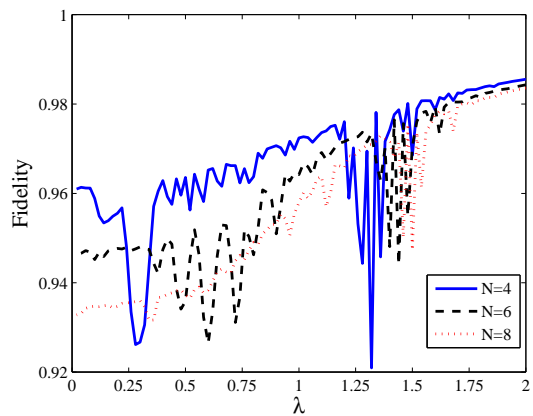}
\caption{(color online) Fidelity vs $\lambda$ for different $N$'s
with $\mu=1$ and $J_A=J_B=0.1$. Blue solid line for $N=4$, black
dashed line for $N=6$, red dotted line for $N=8$. } \label{fidelity}
\end{figure}

We notice that in the above calculations, the thermodynamical limit
condition is not used. As $\lambda$ varies from 0 to 2, the single
particle energy spectrum goes to continuum as $N$ increases. Since
Fr\"{o}hlich transformation is equivalent to the second order
perturbation theory, it may not be valid to apply it to obtain
effective interaction between the two spins. Therefore, we resort to
mixed-state fidelity to demonstrate our approximation used in this
paper. The mixed-state fidelity is given as \cite{Uhlmann,Jozsa}
\begin{equation}
F(\rho_0,\rho_1)=\text{tr}\sqrt{\rho_1^{1/2}\rho_0\rho_1^{1/2}}\text{,}
\end{equation}
which measures the degree of distinguishability between the two
quantum states $\rho_0$ and $\rho_1$. It has already been applied to
the research on QPT \cite{Zanardi}. Starting from the original
Hamiltonian (\ref{originalH}), we obtain the reduced density matrix
$\rho_0$ for the ground state of the two spins by first
diagonalizing $H$ and then tracing over all the degrees of the Ising
chain. On the other hand, we can also obtain the density matrix
$\rho_1$ of the ground state from the effective Hamiltonian
(\ref{Heff}) of the two spins. In Fig.\ref{fidelity}, the relation
between Fidelity and $\lambda$ is plotted for different $N$'s.
Despite oscillations in some paramter intervals, the numerical
method shows high fidelity over the whole interval. Moreover, curves
of different $N$'s converge as $\lambda$ goes larger.

\section{Conclusion and Remark}

In summary, we have studied the dynamical process of two external
spins simultaneously coupling to a transverse field Ising chain.
With Fr\"{o}hlich transformation, we have deduced the effective
Hamiltonian between these two spins. The matrix elements of the
Hamiltonian are expressed in terms of the DSF, which can be
numerically calculated. Through the numerical simulation, it is
shown that the induced entanglement is enhanced by QPT. And the
effective coupling constants reach maximum near the critical point.
By virtue of mixed-state fidelity, we demonstrate the validity of
Fr\"{o}hlich transformation and numerical simulation. Thus, the
measure of the two spin entanglement can be an illustration of QPT.

Besides the enhancement of the coupling intensity around the
critical point, there are oscillations elsewhere. We remark that QPT
takes place in the thermodynamical limit, i.e.,
$N\rightarrow\infty$. In this limit, the energy spectrum of TFIM
goes to continuum from zero to infinity at some parameter, i.e.,
$\lambda=1$. Therefore, the eigen energy of the external spins will
definitely be resonant with one of the eigen energies of the Ising
chain. In this circumstance, our method may not work well. We notice
that a numerical method, the time-dependent density matrix
renormalization group ($t$-DMRG) \cite{tDMRG}, was applied to
central spin models, which were quite similar to ours. Therefore, in
the near future, we will apply this method to our model to obtain a
better result. We expect smooth curves similar to Fig.1 in
Ref.\cite{Osterloh}.

\section*{Acknowledgement}

We thank O. Derzhko and T. Krokhmalskii for helpful discussions and
kind-hearted suggestions about the powerful method for the
calculation of Pfaffian. This work is supported by the National
Fundamental Research Program Grant No. 2006CB921106, China National
Natural Science Foundation Grant Nos. 10325521, 60635040.


\begin{thebibliography}{99}
\bibitem{note1} More rigorously, Eqs.(\ref{coeffecients}) are correct
on condition that $E_0-E_m\pm\mu\neq0$ for all $E_m$'s. This
requirement can not be fulfilled for all $\lambda$'s in the
thermodynamical limit, e.g., $\lambda=1$ and $N\rightarrow\infty$.
However, these equations are correct in the finite $N$ case with
appropriate $\mu$.
\bibitem{note2} It is different from (\ref{HE}) in that $H_{E}'$ is a TFIM with free ends while $H_{E}$ describes the same model
with a cyclic end. Although there are some minor differences between
these two models, they exhibit the same characteristics of quantum
statistical physics in the thermodynamical limit.
\end{thebibliography}
\end{document}